\documentclass[aps,prl,twocolumn,showpacs,superscriptaddress,groupedaddress]{revtex4}
\usepackage{amsmath, amsthm, amssymb}
\usepackage{graphicx}
\usepackage{latexsym}
\usepackage{amsfonts}
\usepackage{graphicx}
\usepackage{amsbsy}
\usepackage{float}
\usepackage{amsmath}

\begin{document}

\title{Predicting Lifetime of Dynamical Networks Experiencing \\  Persistent Random Attacks}

\author{B.~Podobnik}

\affiliation{Faculty of Civil Engineering, 
University of Rijeka, 51000 Rijeka,  Croatia}
\affiliation{Department of Physics, Boston University, Boston, MA 02215, USA}
\affiliation{Zagreb School of Economics and Management, 10000 Zagreb, 
 Croatia}
\affiliation{Faculty of Economics, 
University of Ljubljana, 1000 Ljubljana,  Slovenia}

\author{T. Lipic}
\affiliation{Rudjer Boskovic Institute, 10000 Zagreb,   Croatia}

\author{D. Horvatic}
\affiliation{Physics Department, University of Zagreb,  
 Croatia}

\author{A. Majdandzic}
\affiliation{Department of Physics, Boston University, Boston, MA 02215, USA}

\author{S. Bishop}
\affiliation{Department of Mathematics, UCL, Gower Street,
London,WC1E 6BT, UK}

\author{H. E. Stanley}
\affiliation{Department of Physics, Boston University, Boston, MA 02215, USA}

\begin{abstract} 

{\bf Empirical estimation of critical points at which complex systems
abruptly flip from one state to another is among the remaining
challenges in network science. However, due to the stochastic nature of
critical transitions it is widely believed that critical points are
difficult to estimate, and it is even more difficult, if not impossible,
to predict the time  such transitions occur
\cite{Nature12,Barnosky,Boettiger,Dai}.  We analyze a class of decaying
dynamical networks experiencing persistent attacks in which the
magnitude of the attack is quantified by the probability of an internal
failure, and there is some chance that an internal failure will be
permanent. When the fraction of active neighbors declines to a critical
threshold, cascading failures trigger a network breakdown.  For this
class of network we find both numerically and analytically that the
time  to  the network breakdown, equivalent to the network lifetime, is
inversely dependent upon the magnitude of the attack and logarithmically
dependent on the threshold.  We analyze how permanent attacks affect
dynamical network robustness and use the network lifetime as a measure
of dynamical network robustness offering new methodological insight into system dynamics. 
   }

\end{abstract}

\pacs{89.75.Hc, 64.60.ah, 05.10.-a,05.40.-a }

\maketitle

Despite being largely robust to outside attacks and 
spontaneous fluctuations, during which a fraction of nodes and links
can become either temporarily or permanently dysfunctional
\cite{Albert00,Cohen00,Gao2011}, most real-world complex systems
ultimately collapse and thus have a finite lifetime. The collapse
commonly occurs when, at a critical point, the output of the system
reaches a critical threshold and the system abruptly shifts from one
phase to another \cite{May77,Nature08,Das}. Examples of real-world
complex systems that experience sudden collapse are numerous, e.g., the
spread of disease in living organisms \cite{Vespignani01}, the spread of
political dissent in a human society, or the spread of a product sales
pattern in economics.

In recent years, network science has been utilized to describe complex
systems
\cite{Watts,Barabasi99,Adamic,Bhalla,Krapivski,Dorogovtsev00,AlbertRMP,Montoya,Buldyrev10,Brummitt},
and many dynamical and temporal networks have been proposed
\cite{Jain4,Vespignani12,Holme12,Saavedra2,Barabasi13,NP13,Antonio14}
to explain the complex dynamics that occur in real-world networks
\cite{Vespignani01}. However, the large majority of studies on network science
 analysed separately  either (a) network robustness \cite{Albert00,Cohen00,Gao2011}
   or (b) empirical indicators of critical transitions \cite{Nature12,Barnosky,Boettiger,Dai,Nature08}. 
  The latter group of studies focuses on a phenomenon known as 
   critical slowing down, where as the system approaches a critical point, 
    correlations and variance of the fluctuations increase. 
  Here for (a) and (b)  we propose a common theoretical framework  
  based  on early warning signals to predict
a  network catastrophe (collapse). We focus on 
  an important class of networks characterized by  nodes which 
  flip between two intrinsic states where nodes have the ability to 
   control the state of their neighbours. Examples for these networks range
    from 
   cascading processes in interdependent networks \cite{Buldyrev10,Brummitt}, 
    epidemic spreading in scale-free networks \cite{Vespignani01}, 
    dynamical networks with spontaneous recovery \cite{Antonio14}
    to the glassy dynamics of 
     kinetically constrained  models \cite{Ritort}.

We find that introducing permanent failure produces networks that are
continuously decaying, but the decline is continuous only up to a
specific, critical point---the network lifetime ($t_c$)---when
 the network abruptly collapses. We
find that this critical point  is logarithmically dependent upon the
threshold and inversely dependent on the size of the outside attack.

{\bf Results.} 
So far, no single network model has been able
to explain  the process of finite-time decline with the possibility
of spontaneous collapse,  how ageing or continuous ongoing
time-dependent attacks affects dynamical network robustness, and  how
long a network can continue to function before it collapses or how to
predict the time of network collapse ($t_c$) prior to its occurrence.
To address these three issues associated with dynamical networks
experiencing ongoing time-dependent attacks, we propose a decaying
mechanism in the evolution of a network.  We base this mechanism on
three assumptions.

(i) The function of a node  $n_i$ is dependent upon its nearest neighbors.
 Generally speaking, a network is robust if its nodes are able to function even
  with a large fraction of failed nearest neighbors,  
here denoted by $T_h$. If at time $t$ the fraction of the active neighbors of
node $n_i$ is smaller than or equal to $T_h$, at time
 $t+1$ node $n_i$ will become
externally inactive with a probability $r$.  We use a fractional
threshold \cite{Pod14,EPL14}, which is more appropriate than the absolute
threshold \cite{Watts2,Antonio14} when networks have heterogeneous
degrees.

(ii) Each node can internally fail independent of other nodes, with a
probability $p$ quantifying the magnitude of the external attack.  This
dynamical case is equivalent to the static network case when robustness
is studied under simultaneous random or targeted attack
\cite{Albert00,Cohen00,Holme02,Song06,Gao2011,Holme12}.

(iii) Although we assume that a node can recover from an internal
failure after a finite period of time ($\tau$)  has passed, because
internal failure always carries the potential risk of being permanent,
by   $1 - q$ we define such an event.  Thus, due to  ongoing time-dependent attacks, when a
network is attacked not every node will be able to recover. As 
in Ref.~\cite{Antonio14}, a node can be considered active only if it is
both internally and externally active.
For a more complicated network, that includes adding new nodes 
and introduction of link failures,   see Methods and Model extensions.

Note that the dynamical approach used to quantify the robustness of a
network under time-dependent attacks can also be used to address the
ageing process. In this case the probability $p$ serves as an internal
network property \cite{Antonio14}.  Over time not every
internally-failed node can recover, e.g., at the cell level in biology
there may be a defective apoptotic process \cite{Karam}.

\begin{figure}[b]
\centering \includegraphics[width=0.4\textwidth]{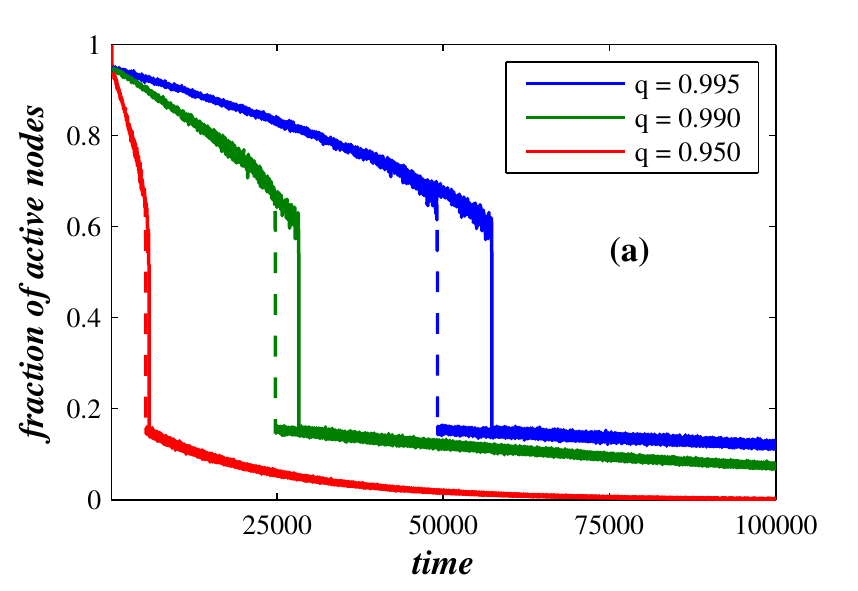}
\centering \includegraphics[width=0.4\textwidth]{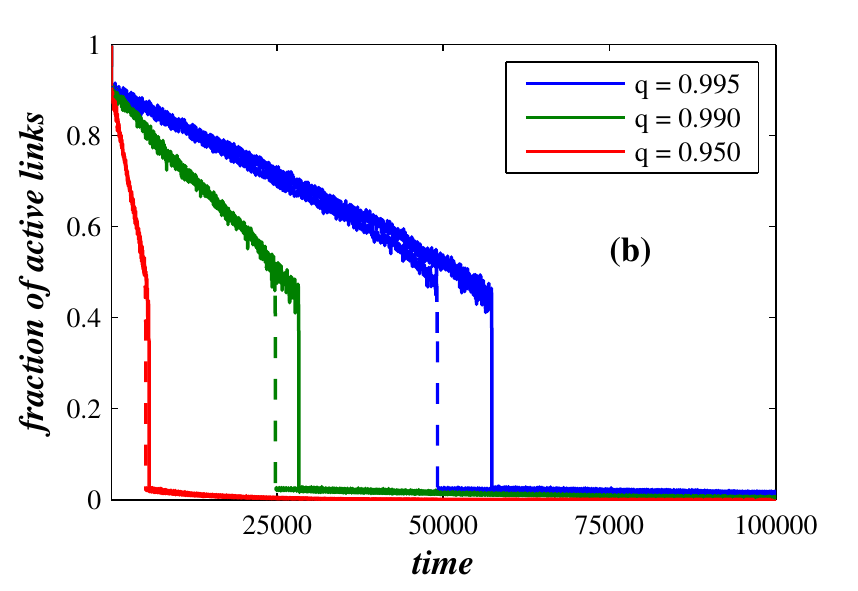}
\centering \includegraphics[width=0.4\textwidth]{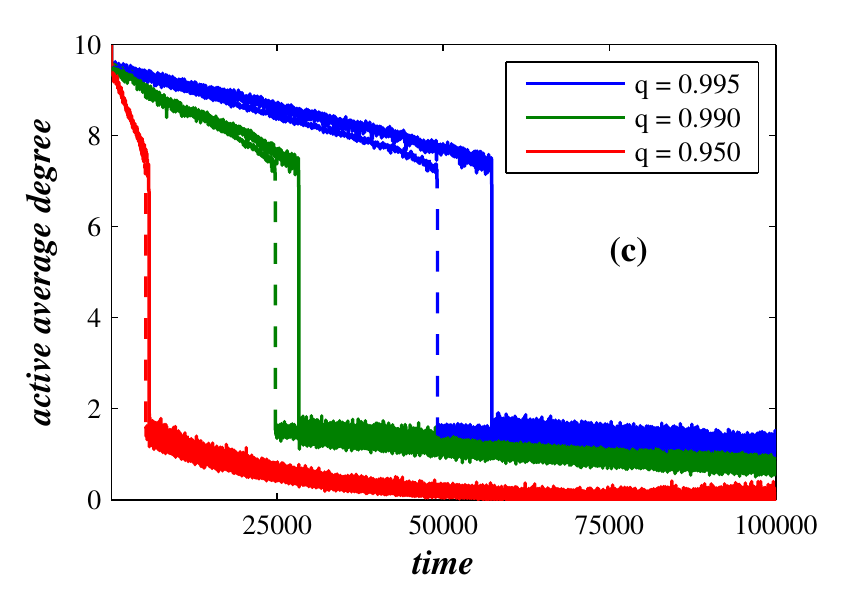}

\caption{Comparison between BA (solid line) and ER (dashed line)
  decaying dynamical networks, when $\langle k \rangle = 10$. Under the
  same level of attack, quantified by equal $p$, the BA
  decaying network exhibits the higher level of dynamical robustness
  than the ER decaying network, where dynamical robustness is quantified
  here by larger $t_c$. For (a) the fractions of active nodes and (b) links and
 (c) the average degree as a function of time the BA decaying networks are
  more robust than the ER decline networks. Here the robustness is
  defined in a dynamical way---the more robust  a network is, the longer it
   lasts. For both networks we use $p = 0.001$, $r = 0.8$, $T_h = 0.5$,
  and $\tau = 50$.  }
\label{2}
\end{figure}

\begin{figure}[b]
\centering \includegraphics[width=0.4\textwidth]{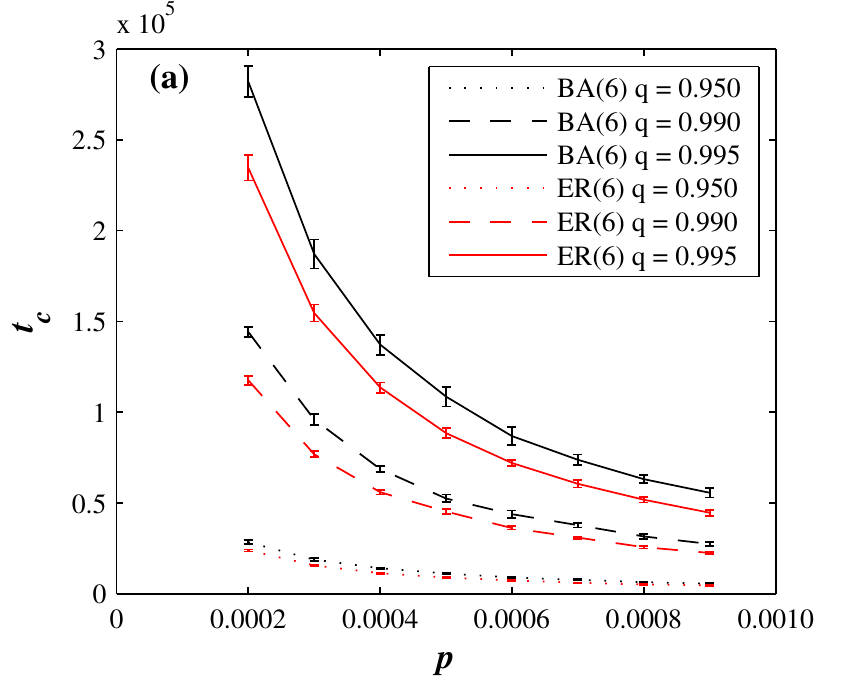}
\centering \includegraphics[width=0.4\textwidth]{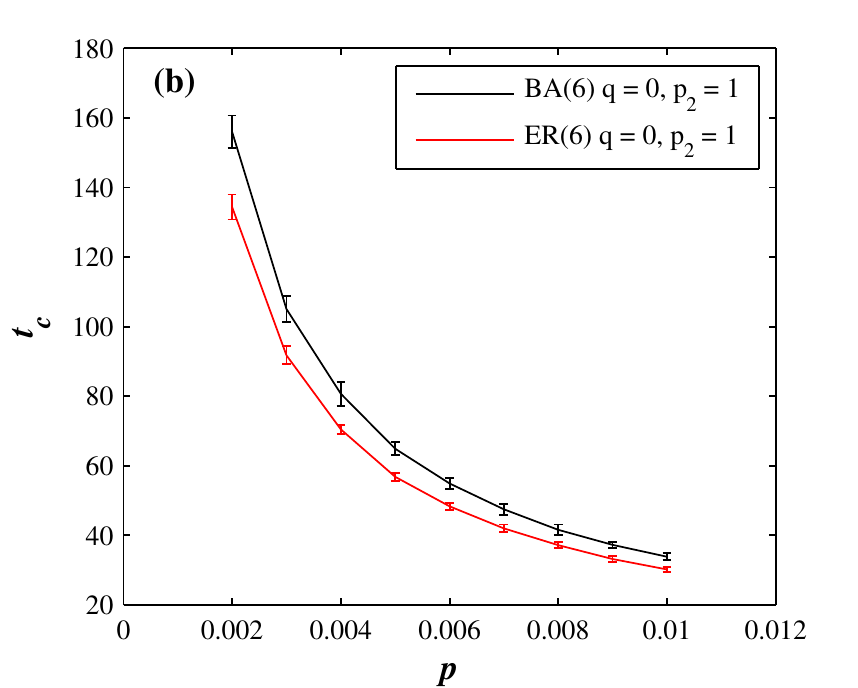}
\caption{Dynamical robustness of BA and  ER decaying networks 
   for different values of $q$, where $t_c$ serves as a measure
  of dynamical robustness. We show  $t_c$ versus $p$ for (a) $q=0.995$,
  $0.95 - 0.995$.  The BA decaying networks
  exhibit higher level of robustness than the ER decaying networks. We
  use $r = 0.8$ and $\langle k \rangle$ = 6.  (b) For BA and ER
  decaying dynamical networks when a node cannot recover ($q=0$), for
  $\langle k \rangle = 6$ the larger time $t_c$ (the larger dynamical
  robustness) we obtain for BA decaying network than for ER decaying
  network. Here, the larger time $t_c$, the longer the network lasts, the more robust the network.  }
\label{3a}
\end{figure}

In seminal studies of the stability of large complex systems, Gardner
and Ashby \cite{Nature70} (numerically) and May \cite{MayNature72}
(analytically) examined how they maintain stability up to some critical
level of connectedness and then suddenly, as the level of connectedness
increases, become unstable. In contrast, our focus is on the reverse
process in which a continuous decrease in the number of functioning
nodes and links between them, over time decreases network complexity.  We describe the
level of network functionality in terms of the fraction of the total
number of  nodes that are continuing to function \cite{Antonio14}
and the fraction of active links, $f_l$ \cite{Pod14}.

To determine how $T_h$ and $q$ affect network functionality, we first
generate three different random  Erd\H{o}s Renyi (ER) networks
 and scale-free networks (Barabasi-Albert BA model), each
with initial $N(t=0)=10,000$ nodes and an average degree $\langle k
\rangle = 10$.  Then each network is placed under persistent attack,
quantified by $p$, where we allow nodes in the inactive phase to
permanently fail with a probability $1 - q$.  For varying values of $q$
for each decaying  network, Fig.~\ref{2} shows the fraction of active
nodes, the fraction of active links, and the average degree, each as a
function of time.  Figure~\ref{2} shows that when the threshold is fixed
but the values of $q$ varied, the network decays continuously until at
some specific time $t_c$ it abruptly and spontaneously collapses, a
collapse brought on by a sudden large increase in the number of
 inactive links and nodes. Note that this sudden collapse
does not require any parameter other than $q$. As expected, the more
rapid the exponential decay, the sooner the network will collapse. 
 When we compare the fractions of active nodes and
links we see that, when the network collapse occurs, the fraction of
active links decreases substantially more than the fraction of active
nodes, where the first fraction approximately equals $1 - r$.

\begin{figure}[b]
\centering \includegraphics[width=0.4\textwidth]{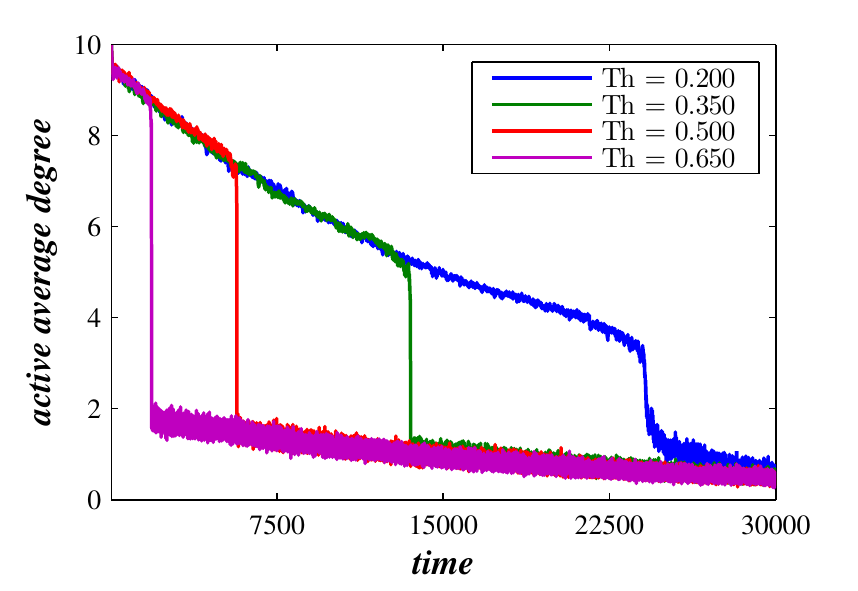}

\caption{ Time of network crash $t_c$ versus  the fraction of active links for different values of the 
network threshold, $T_h$ calculated 
  for BA
  dynamical network.  $t_c$  dramatically depends on $T_h$. } 
\label{3}
\end{figure}
 
Note that our analysis addresses both how long a dynamical network will
function before collapse and how robust it will be under long-term
continuous attack, i.e., the larger the network lifetime $t_c$,
the more robust the network.  The study of robustness of dynamic
networks under continuous attack is highly relevant to the concerns of
both researchers and practitioners.  For instance, important goal of military
science is determining how a military network can remain robust under
persistent enemy attack; or a goal in finance is determining how a financial
system can remain robust when a fraction of its banks fails over time.
References~\cite{Albert00,Cohen00} reported that scale-free networks are
more robust than ER networks to multiple random and simultaneous attacks.
Next we compare the robustness of decaying  BA and ER networks.

 Here the network robustness is
  defined in a dynamical way, where the more robust  a network is,
  the longer it lasts. 
Figure~\ref{2} compares the  dynamical robustness of decaying  BA and ER
networks under continuous long-term attack, measured by $p$, as a
function of time. We use the same parameters for both types of networks.
Using network lifetime to quantify dynamical network robustness,
Fig.~\ref{2} shows that, even when nodes and links fail, BA networks are
generally more  robust than ER networks.  A dynamical BA network is
typically able to survive with higher values of $t_c$ than an ER network
before it exhibits an abrupt drop in the fraction of its active nodes
and links and its average degree.  Figure~\ref{2}(c) shows that for
 the $q$ values displayed, the average degree of a dynamical BA network
lasts longer (i.e., $t_c$ is larger) than the average degree of a
dynamical ER network.

\begin{figure}[b]
\centering \includegraphics[width=0.4\textwidth]{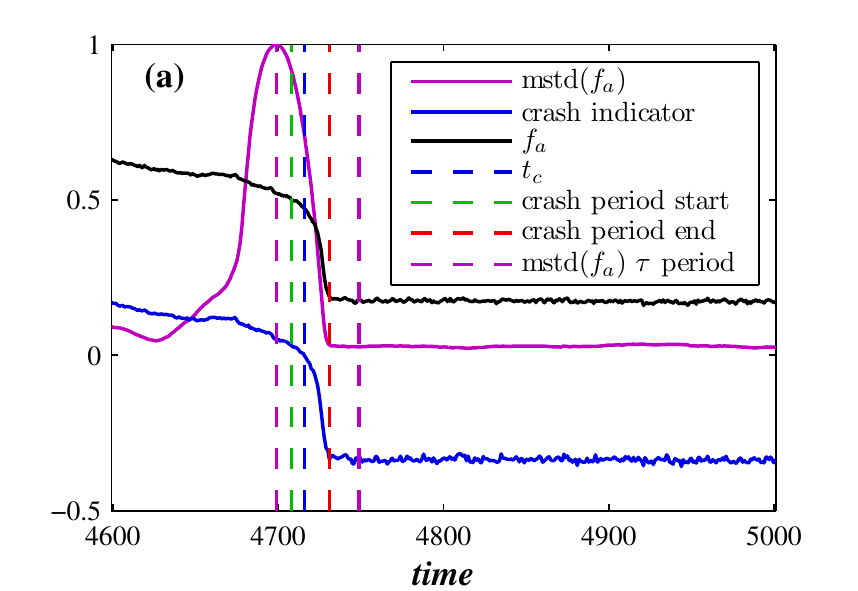}
\centering \includegraphics[width=0.4\textwidth]{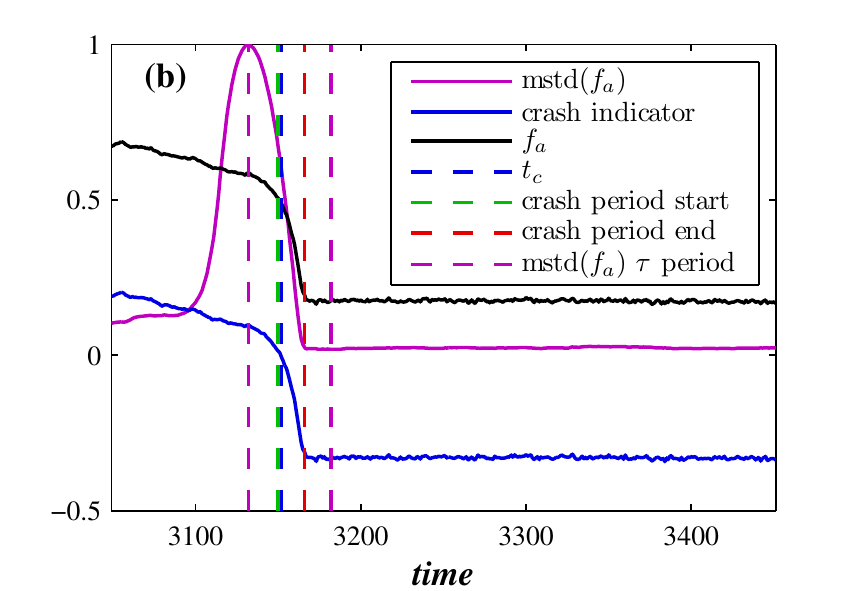}
\caption{Indicators for network crash. Shown are 3 indicators together
  with the fraction of active links obtained for (a) ER and (b) BA
  decaying network when $q = 0.99$. We show indicator I (mstd), moving 
  standard deviation of fraction of active nodes (forward method) with window
   size $\tau = 50$ 
  (the same $\tau$ as in recovery process), and indicator II (crash indicator), the
  average fraction of active neighbors in excess of threshold
  $T_h$. Approximately, when the average  fraction of active neighbors exceeds
 the threshold, cascading failures trigger a
network breakdown. The cascade lasts several decades. 
We also show the analytical   indicator III ($t_c$).  We use $p=0.003$, $T_h
  = 0.5$, and $r = 0.8$.}
\label{4}
\end{figure}

To determine whether this difference in   dynamical robustness is general or
sample-dependent, we compare decaying BA and ER networks for different
$q$ values.  For fixed $T_h$ and $r$, and varying values of 
    $q$, Fig.~\ref{3a}(a) shows the relationship
  between $t_c$, whose precise definition 
   will be explained in Fig.~\ref{4}, and $p$.
    The BA decaying networks
generally exhibit a higher level of  dynamical robustness than the ER decaying
networks. We find similar dependence between   $t_c$ and $p$ for varying 
 network degrees.

The dynamical approach also allows the non-trivial possibility that, 
following an attack, nodes can remain permanently damaged, corresponding
to the $q = 0$ case in which no recovery is allowed. This case is
important because it allows us to compare our results with studies in
which robustness is analysed in static networks under either
simultaneous random or targeted attack
\cite{Albert00,Cohen00,Holme02,Song06,Gao2011,Holme12}.  Note that when
$q = 0$, the period $\tau$ becomes irrelevant. For a fixed $T_h=0.5$ and
$\langle k \rangle = 6$, Fig.~\ref{3a}(b) shows how the time $t_c$  is
affected by the level of outside attack, quantified by $p$.  Again as in
Figure~\ref{3a}(a), the larger the level of outside attack, the smaller the
network lifetime.  Because no recovery is allowed for this case, the
lifetime values in Fig.~\ref{3a}(b) are much smaller than in
Fig.~\ref{3a}(a).  When $q = 0$, decaying BA networks exhibit a
significantly higher level of dynamical robustness than decaying ER networks, and
the times $t_c$  calculated for BA networks are consistently larger than
the times  $t_c$  calculated for ER networks---the dependence between $t_c$ and
  $p$ follows a
hyperbolic function. 

\begin{figure}[b]
\centering 
\centering \includegraphics[width=0.4\textwidth]{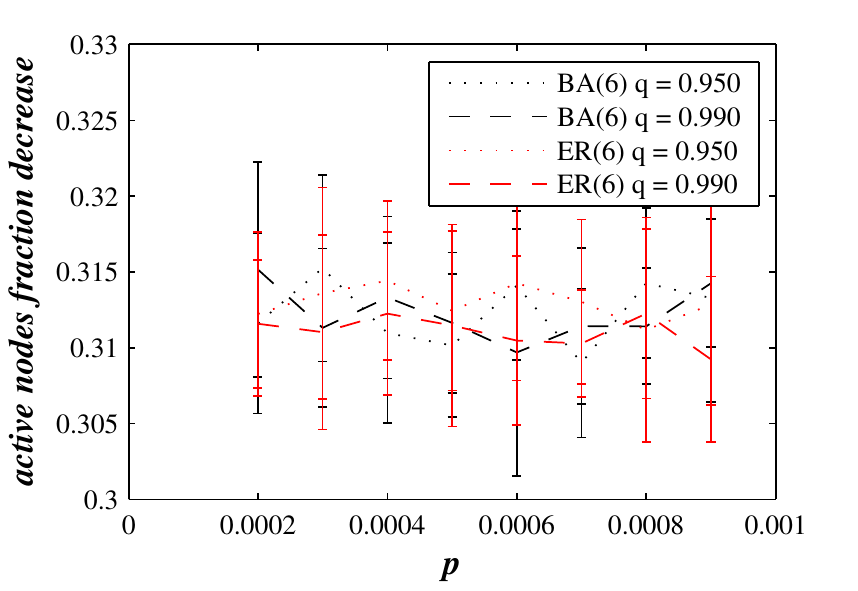}

\caption{ For the varying $q$ values, for both decaying ER and  
decaying BA network we show the
  size of the decrease of $f_n$ during the network crash, 
  equal to $T_h  - r$.}
\label{5}
\end{figure}

\begin{figure}[b]
\centering \includegraphics[width=0.4\textwidth]{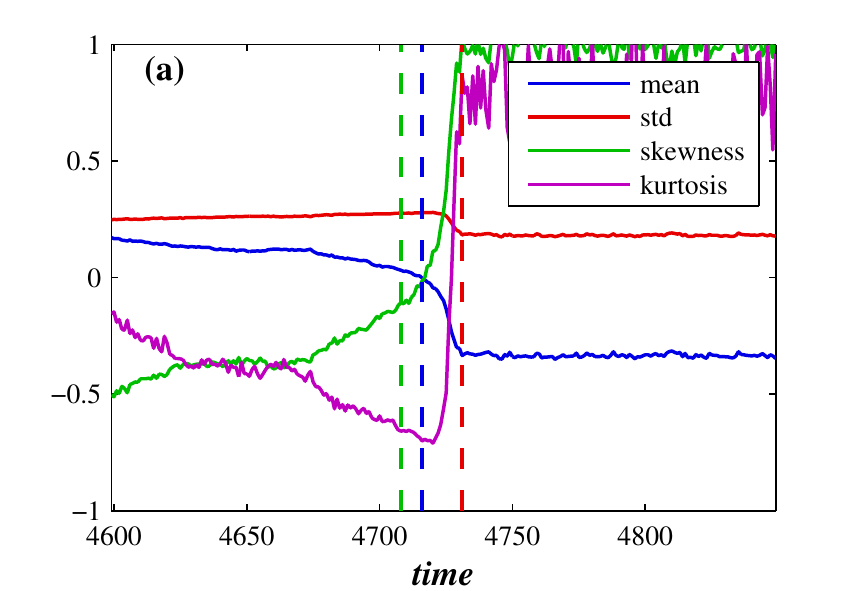}
\centering \includegraphics[width=0.4\textwidth]{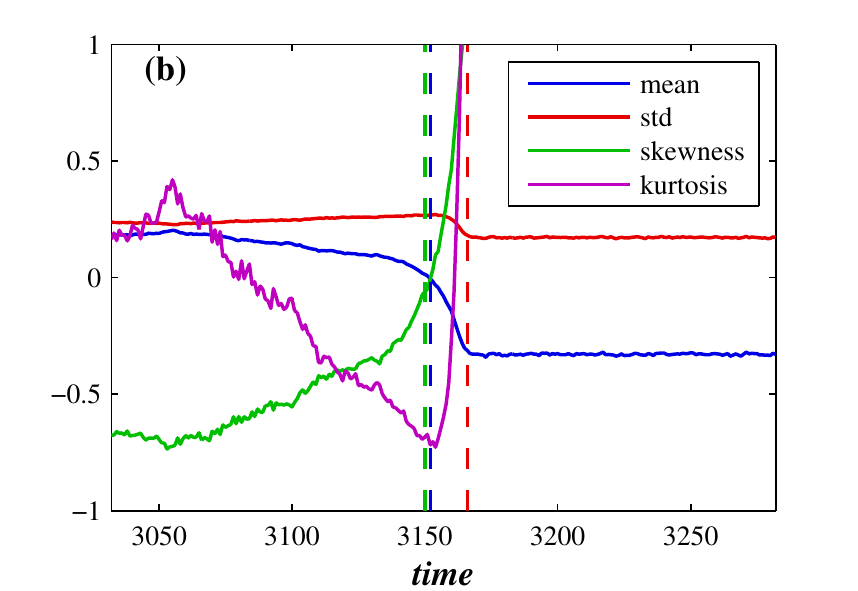}

\caption{Statistics of indicator III for (a) decaying ER and (b) 
decaying BA network.
 We show how the first four moments
  of indicator III change over time just before, during and just after
  the crash. Just before the network crash, skewness and kurtosis
  dramatically and abruptly increase. 
 }
\label{6}
\end{figure}

\begin{figure}[b]
\centering \includegraphics[width=0.4\textwidth]{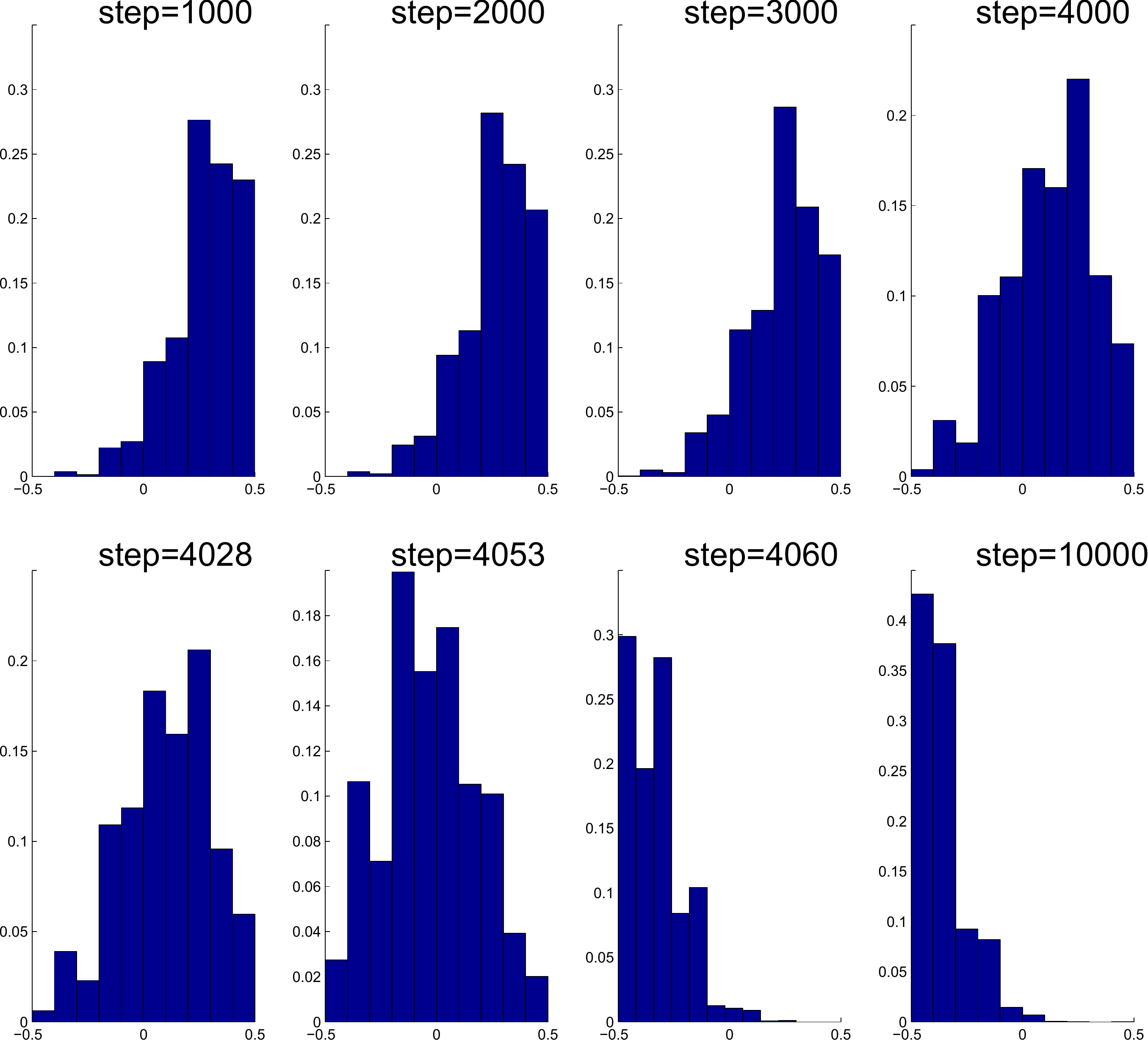}

\caption{Distribution of 
   the fraction of
active neighbors of each node above the threshold $T_h$ substantially changes
 during the network failure. 
 }
\label{66}
\end{figure}

In the decaying network approach, as previously stated,
 we quantify the network threshold 
in terms of the
fraction of failed nearest neighbors a node can sustain and still work
properly. If a node $n_i$ is initially linked to 10 nodes and needs at
least 6 active neighbors to function, then as the network decays node $n_i$
will die when the number of its active neighbors drops to 5.
Figure~\ref{3} shows a decaying BA  network with fixed $q$ and
$p$.  The network lifetime $t_c$ is strongly dependent upon
$T_h$, e.g., when $T_h$ is decreased from 0.5 to 0.2, $t_c$ changes by a
factor of 10.

In a complex dynamical system, thresholds and critical points control the
transition process between different states 
\cite{MayNature}.  A phase transition occurs when a system reaches a system
threshold.  Using this perspective, the network decline and collapse
numerically described above can also be treated analytically.  Note that
at time $t-\tau$, due to attacks, the network has $N(t- \tau)$
living nodes, both active and inactive, among which $p N(t- \tau)$ nodes
switch to the internally inactive phase. 
 The change in the number of living nodes is proportional to the
probability that a node which internally failed at $t- \tau$ 
will not recover,
$1 - q$, thus
\begin{equation}
  d N(t- \tau) =  - (1- q) p   N(t- \tau) dt. 
  \label{exp}
\end{equation}
The number of living nodes remaining is determined by an exponential
decay $N(t) = N(0) \exp(- (1- q) p t)$ with a decay rate of $\lambda =
(1- q) p $, where $N(t)/N(0)$ is the average fraction of living nodes
equivalent to the average fraction of living neighbors of each node. For
simplicity, let us assume that all nodes have the same initial degree
$k$, remembering that node $n_i$ will be active if the number of its
functioning neighbors is larger than $m = T_h k$.  At time $t$, the
fraction of $n_i'$s failed neighboring nodes $N_f(t) / N(t)$ among
$N(t)$ living nodes previously estimated as $k(t) = k \exp(- (1- q) p
t)$, can be approximated using the probabilities of internal and
external failures, $p$ and $E[m,k(t)]$,
\begin{equation}
  a \equiv a(p,r,m,k(t)) = p + r (1 - p) E[m,k(t)],
  \label{a}
\end{equation}  
where $E[m,k(t)] = \Sigma_{j=0}^m {k(t) \choose k(t)-j} a^{k(t)-j}
(1-a)^{j}$. In the above figures we analysed the ratio between the
living active and the initial number of nodes $N_a(t) / N(0)$, which is
equivalent to
\begin{equation}
  f_n \equiv  \frac{N_a(t)}{N(t)} \frac{N(t)}{N(0)} = (1-a) \exp[- (1- q) p t]. 
\label{indicator}
\end{equation}  
Here $\exp[- (1- q) p t]$ is due to the continuous decline of the
network, and $1-a$ indicates the sudden network crash occurring when the
fraction of living nodes $k(t)/k$ approaches the threshold $T_h=m/k$.
At this limit, from Eqs.~(\ref{a}) and (\ref{indicator}) we see that
$E[m,k(t)] \rightarrow 1 $ and the probability of external failure
dominates $p$. 

Although predictive power is important in any scientific endeavour, it is
widely assumed that predicting critical transitions is extremely
difficult because any change in the dynamic state of a system
immediately prior to reaching the critical point will be slight
\cite{Nature08}.  Unlike early-warning indicators that utilize recent
system outputs to detect impending system collapse, we use past data to
estimate network parameters, from which we generate numerical
simulations that describe the network state at any future time,
including the moment of network failure $t_c$.  Thus the dynamical approach
allows us to precisely predict the network crash.

Figure~\ref{4} shows the predictive power of this dynamical approach when
applied to both (a) ER and (b) BA networks in the process of decaying.  It
shows the fraction of active nodes for time scales that include the
network crash and also two important values: (I) the standard deviation
of the fraction of active nodes calculated using moving boxes similar to
the approach proposed in Ref.~\cite{Nature08} and (II) the average fraction
of active neighbors---active both internally and externally---above the
threshold $T_h$. Equations~(2) and (3) provide a theoretical explanation
for II, predicting that when the fraction of
currently living nodes approaches the threshold that controls the
external failures, the probability  of critical transition increases
 and finally cascading failures lasting several decades trigger a
network breakdown. 
Note that just prior to network crash, indicator I
increases substantially and thus is an indicator of impending failure
and a predictor of the time $t_c$, i.e., it predicts when network failure
will occur. Reference~\cite{Nature08} suggests that as a dynamical
system approaches a critical threshold its state at any given moment
increasingly resembles its previous state, implying an increase in
variance and autocorrelation as reported for dynamical networks in Ref.~\cite{Pod14}.  
 We find that this predictive power holds
for both BA and ER decaying networks  for a wide variety of
parameters. In Fig.~\ref{4} the values of II shift from positive to
negative and the time when  this occurs can also be used as a predictor of
the time of network collapse $t_c$.  In addition to the two numerical
indicators, we also suggest a third analytical indicator (III), obtained
by equating $\exp[- (1- q) p t]$ in Eq.~(\ref{indicator}) with $T_h$,
\begin{equation}
   t_c = - \ln(T_h)/p(1 - q).
  \label{tc}
\end{equation}   

In the decaying network approach, the proportional threshold $T_h$ controls
both node-level crashes and ``macro'' network-level crashes, which makes
$T_h$ a breakdown threshold.  For a general set of network parameters,
the time of network failure is the ratio between the closeness of the
fraction of active nodes to the threshold, $- \ln(T_h)$ and the rapidity
of the approach of the fraction to the threshold $p(1 - q)$.  In
Figs.~\ref{4}(a) and \ref{4}(b) we see that $t_c$ typically corresponds
to the beginning of the network failure. Additionally, for varying values
of $q$ in Fig.~\ref{5}, for both decaying BA and ER networks, we find
how much the fraction of active nodes decreases during the network
crash. It decreases for $T_h - r$.
  
Figures~(1) through (4) show that Eq.~(\ref{tc}) agrees well with the
numerical results that describe how $t_c$ is affected by $T_h$, $ p$,
and $q$.   
 Figures~\ref{4}(a) and \ref{4}(b) show that the values of
indicators II and III are nearly identical, suggesting that during a
network crash the time when the fraction of living nodes approaches the
threshold is approximately identical to the time when the fraction of
active neighboring nodes approaches the threshold, implying that in
Eq.~(\ref{indicator}) the second term contributes significantly more
than the first. Further we find that the time series of the fraction of
active neighbors of each node above the threshold $T_h$ substantially
changes over time.  For this time series, calculated for each network in
Figs.~\ref{4}(a) and \ref{4}(b)  in addition to the average
representing our indicator III, Fig.~\ref{6}(a)-(b) shows the higher moments,
variance, skewness, and kurtosis. For  different times
 prior, during,  and after the network failure,
  Fig.~\ref{66} shows the distribution 
 of the fraction of
active neighbors of each node above the threshold $T_h$. 
 Just prior to network failure,
skewness and kurtosis and the distribution  dramatically change.   
      
In estimating the time of a future network crash, we calculate network
parameters from its creation at time $t=0$ to some time $t$ during which
the fraction of active nodes $f_n$ decreases from 1 to some value $f'$.
When we numerically generate the network we thus record the time $t$ at
which $f_n$ reaches $f'$.  Then the true prediction of future network
collapse is not $t_c$ but $t_c - t$, i.e., $t_c$ represents the entire
lifetime of the network, and our interest is in estimating the lifetime
remaining after $t$.

{\bf Conclusion.}
In both the natural and social sciences a wide range of
real-world complex networks have a finite lifetime characterized by
abrupt shifts between phases.  One of the differences between the
behaviour of natural systems and social systems near a critical point is
that the sudden random change commonly occurs at a fixed point in
natural systems \cite{Lebowitz}. In contrast, in social sciences even a
critical point is a random variable and, although there is an
expectation that some radical change is imminently probable, it is
generally assumed that it is not possible to predict the exact time of its
occurrence.  We have explored the predictive power  for a particular class
 of dynamic non-equilibrium decaying  networks.  

 In our network model, the
same parameter, a proportional threshold $T_h$ responsible for
node-level crashes, also controls ``macro'' network-level crashes. 
 Our dynamical network study includes cases in which nodes can
recover $(q \ne 0)$ or remain in a failed state $(q = 0)$ after a
failure caused by the ageing process or by severe hostile time-dependent
attacks.  Model extensions are included in the Methods section. 
How successful  our network might be when applied in practice 
 depends first, on how capable we are to estimate the model parameters, especially
  the critical threshold \cite{Zhao}, and  
  second, how good  our network is as a proxy for  
  a real-world complex system.  In practice the second issue is very frequent 
  when real-world stochastic systems composed of large number of units, such as financial,  are modelled with theoretical 
  stochastic processes, 
   such as multivariate autoregressive for instance. Besides, real-world networks 
   rather interact with other networks 
   \cite{Buldyrev10,Brummitt} than act as a single network \cite{Binder}. 
   Due to interdependencies between different networks, a shock on a 
    particular network may change not only its own network parameters,
     but also
     the  parameters of all other networks. 
    Clearly, since the
time  to  the network breakdown depends on network parameters, 
    any change in network parameters also  affects our estimation   
 on the timing of network collapse. 

{\bf Methods and Model Extensions}    
   
\begin{figure}[b]
\centering \includegraphics[width=0.47\textwidth]{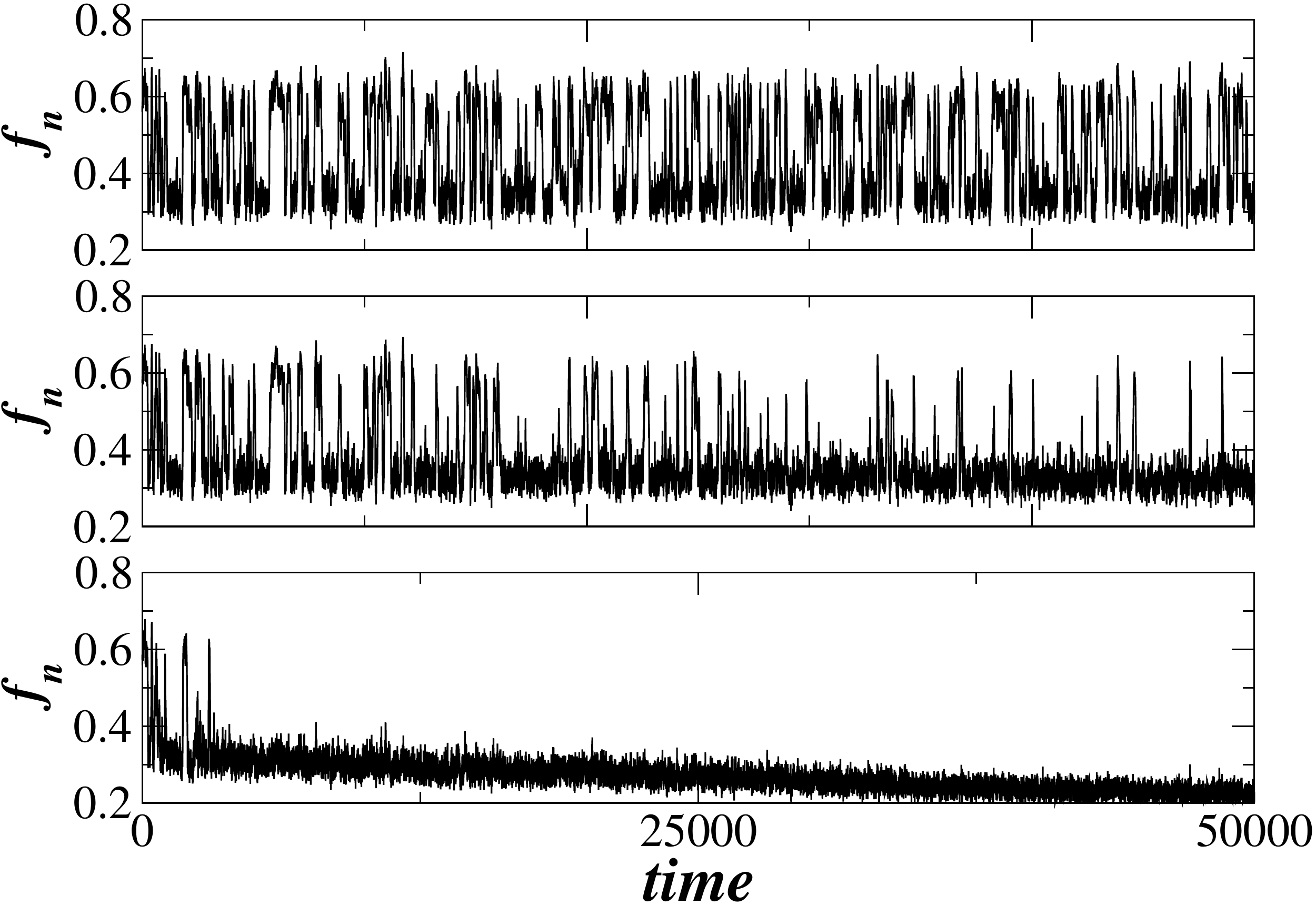}
\caption
{ Phase-flipping with decay mechanism for the decaying
 BA network
 with parameters
  $T_h = 0.50$,  $\tau = 50$, the average degree 
  $\langle k \rangle = 3$, 1000 nodes. 
 With decreasing $q$, from $0.999995$, $0.9999$, and $0.999$ (from top
  to bottom)  we show
 that the average $f_n$ decays  with time where the smaller
  $q$  the faster the decay.   
}
\label{decay}
\end{figure}

    {\bf  Phase-flipping with decay} 
   Ref. \cite{Antonio14} reported that introduction 
   of both dynamical recovery and
stochastic contiguous spreading leads to spontaneous 
collective network phase-flipping phenomena. For
 $q=1$ case, 
 we choose the network parameter set to  enable phase-flipping
  between two stable states. 
 When decay mechanism is included ($q \ne 1$),  in
Fig. \ref{decay} 
 with decreasing parameter $q$, the fraction of living nodes
gradually disappears over time and so the phase-flipping 
  phenomena and collective network mode.

\begin{figure}[b]
\centering \includegraphics{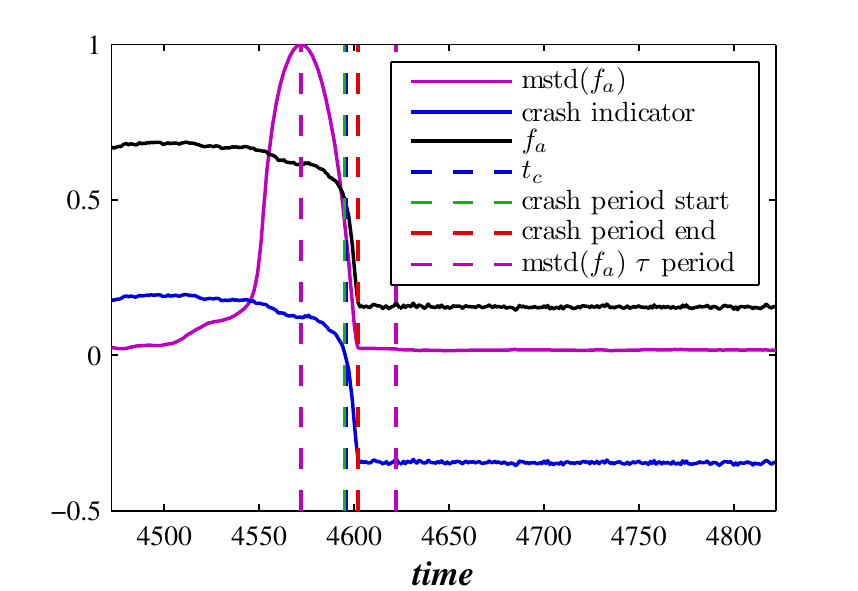}
\caption
{Indicators for network crash, similar as in Fig.4 but now we include
 internal link failures.  Shown are 3 indicators  obtained for an ER 
  decaying network when $q = 0.99$, $T_h = 0.99$, $r=0.8$, $p=0.003$, and
   $p_{\ell}=0.003$.     
}
\label{methods}
\end{figure}

  {\bf  Introduction of link failures.} 
   For decaying networks,
    it is reasonably to assume  that  links 
   as well as  nodes may also fail at any time $t$. 
 Thus,  we additionally 
  consider that every link $\ell$ can independently fail 
    with  probability $p_{\ell}$ (internal failures in links). 
       We define that a link is ``healthy'' (active)  if it is active both internally, controlled by $p_{\ell}$, and externally, controlled 
        by  $p$ in (ii). 
 When a dynamical network is defined one should find out which part of 
 the parameter space is characteristic for  a stable network regime and 
  which part of the space determines an unstable network regime.  
 For our dynamical network defined by three probability parameters, 
 $p$,   $r$,  and $p_{\ell}$, 
  next, for a case when nodes always recover after being
  inactive for a while ($q=1$), we provide   analytical 
  results for the 
 3D hysteresis in parameter space  here  enclosed by manyfolds
comprising spinodals which separate regions of stability and instability.  
   Thus, for an active   node $i$ with $k_i$ neighbors,
    the link between $n_i$ and another node 
   $n_j$ fails either due to an internal link failure  ($X$)
   with probability $p_{\ell}$ or due to  external failure of node $j$ ($Y$) 
    equal to $a$---in the latter case  according to definition, when 
    $n_j$ is failed,  all its links  have failed. Thus,
     assuming $X$ and $Y$ are 
    independent, 
   the link  is active with probability $(1-a)(1-p_{\ell})$
   and accordingly inactive with probability $ 1 -  (1-a)(1-p_{\ell}) =
   a + p_{\ell} - a p_{\ell} \equiv a_{\ell}$. Similar to
    (i), now when even the links can fail internally,
    we define that  node $n_i$  is active only if there are more
    than  $ m$ active links (not nodes as in (i)). Thus by  
\begin{equation}    
     E(k,m,a_{\ell}) \equiv 
  \Sigma_{j=0}^{m} a_{\ell}^{k-j} (1 - a_{\ell})^{j} 
   {k \choose k -  j},  
 \label{E1}
\end{equation}    
    we define  
 the probability that node $n_i$'s neighborhood is critically damaged
  with more than $m$ broken links. Finally, we derive 
   the probability that a randomly chosen  node $n_i$ with degree $k$ 
    is  inactive,  equal to  the fraction of inactive nodes, 
\begin{equation} 
 a_k \equiv a = p + r (1 -p)  E(k,m,a_{\ell}),   
 \label{a1}
\end{equation} 
 where we apply  probability theory  
 $P(X \cup Y) =P(X) + P(Y) - P(X) P(Y)$ for two independent 
  events. Clearly, this equation is derived under the assumption
   of independence of the external and internal failures
    is only approximately true 
   since internal failures affect external failures. 
  For an ER decaying network.
   Fig.~\ref{methods} shows the predictive power for a case when internal link failures
  are considered.

\begin{figure}[b]
\includegraphics[width=0.5\textwidth]{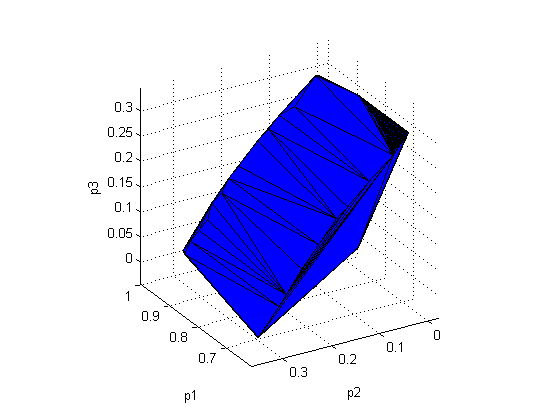}
\caption
{For $q=1$ case,  we show the 3D phase diagram in model parameter space 
$(p_1, p_2,p_3)$
representing hysteresis region.    
}
\label{11}
\end{figure}

 For simplicity  for the moment we analyze random regular networks 
  since Eq.~(\ref{a1}) assumes that all nodes have the same degree. 
  Fig.~\ref{11} shows
the mean-field (MFT) prediction for the 
 position of spinodals in 3D parameter space $(p_1, p_2 \equiv r, p_3\equiv p_{\ell})$ 
 obtained from Eq.~(\ref{a1}), where $p_1 = 1 - \exp(-p \tau)$~\cite{Antonio14}.
   In 3D parameter space,  
 the  3D  hysteresis region is enclosed by planes comprising spinodals.  
 The 3D  space is obtained by fixing $p_{\ell}$
 and calculating  the  2D hysteresis region  bounded with two 
 spinodals  merging at a critical point.

 {\bf Addition of new nodes.}  
  In the growth of a network, the model can be additionally extended
  in a way that nodes can be also created not only destroyed. 
  We can assume that at each time $t$,  among $N(t)$
living nodes,  
 each living node can create a new node
  with   probability $p'$. Then Eq.~(\ref{exp}) is modified to 
   \begin{equation}
  d N(t- \tau) = (p'  - (1- q) p)   N(t- \tau) dt, 
  \label{exp1}
\end{equation}
   and the system can not only exponentially decay, but also increase and even
    stay stable if $p'  = (1- q) p$.

\end{document}